# A memristive deep belief neural network based on silicon synapses

Wei Wang[1,2,*], Loai Danial[1,5], Yang Li[1,2], Eric Herbelin[1], Evgeny Pikhay[3], Yakov Roizin[3], Barak Hoffer[1], Zhongrui Wang[4], Shahar Kvatinsky[1,*]

[1] The Andrew and Erna Viterbi Faculty of Electrical and Computer Engineering, Technion-Israel Institute of Technology, Haifa, Israel 3200003

[2] Peng Cheng Laboratory, Shenzhen, China

[3] Tower Semiconductor, Migdal HaEmek, Israel 2310502

[4] Department of Electrical and Electronic Engineering, The University of Hong Kong, Pokfulam Road, Hong Kong

[5] Current affiliation: Intel Corporation, IDC, Haifa, Israel

[*] Email: wangwei@pcl.ac.cn; shahar@ee.technion.ac.il

**ABSTRACT**

Memristor-based neuromorphic computing could overcome the limitations of traditional von Neumann computing architectures — in which data are shuffled between separate memory and processing units — and improve the performance of deep neural networks. However, this will require accurate synaptic-like device performance, and memristors typically suffer from poor yield and a limited number of reliable conductance states. Here we report floating gate memristive synaptic devices that are fabricated in a commercial complementary metal–oxide–semiconductor (CMOS) process. These silicon synapses offer analogue tunability, high endurance, long retention times, predictable cycling degradation, moderate device-to-device variations, and high yield. They also provide two orders of magnitude higher energy efficiency for multiply–accumulate operations than graphics processing units. We use two 12-by-8 arrays of the memristive devices for in-situ training of a 19-by-8 memristive restricted Boltzmann machine for pattern recognition via a gradient descent algorithm based on contrastive divergence. We then create a memristive deep belief neural network consisting of three memristive restricted Boltzmann machines. We test this on the modified National Institute of Standards and Technology (MNIST) dataset, demonstrating recognition accuracy up to 97.05%.

**Introduction**

Memristive devices operating as artificial synapses could be used to create efficient and low-power neuromorphic computing systems, performing analogue computation in the memory device itself and avoiding some of the bottlenecks associated with traditional computing systems[1–3]. There is, however, a mismatch between the realistic synaptic behavior of the memristive devices and the stringent requirements of neural network algorithms. This means that use of memristors as synaptic weighting elements in mainstream artificial neural network-based hardware remains challenging[4,5].

To be used in hardware based deep neural networks (DNNs), memristive devices need to have high production yields, and offer a large number of conductance states, high uniformity, long retention times and high endurance[6,8,9]. Emerging non-volatile memory devices — including resistive random-access memory (RRAM)[10–13], magnetic random-access memory (MRAM)[14], phase-change memory (PCM)[15,16], ferroelectric memory[17], and electrolyte-gated transistors[18–20] — have been used as synaptic devices in neuromorphic computing. However, few have been able to demonstrate all of the desired requirements[21]. Many of these technologies also have a low production yield[22] or need special materials or processes[23] when being integrated into industrial complementary metal–oxide–semiconductor (CMOS) processes.

Error backpropagation-based gradient descent[24] is a key algorithm used in training DNNs. The errors are usually small and need to be handled by high-precision neural circuits. Additionally, the data dependency of gradient descent on the backpropagated errors often needs excessive memory to store the intermediate neuron states[25,26]. High accuracy training of DNNs using phase-change memory based analogue memristive devices has been demonstrated, although the high precision neural circuits were implemented in software[27]. A deep convolutional neural network (DCNN) has also been implemented in hardware using RRAM-based synaptic weight matrices and CMOS-based neuronal circuits[28]. However, the CMOS-based neurons are complex due to the use of analogue-to-digital and digital-to-analogue converters (DACs and ADCs) and the digital circuit for implementing the activation function in the digital domain. Most of the weights are also pre-trained on a digital computer, rather than being optimized in-situ using the error backpropagation algorithm, which leads to additional hardware overhead.

Spiking neural networks (SNN), which directly resemble biological neural networks, have two advantages for hardware implementations. First, the spiking signal as a binary input and output simplifies the neural circuits. Second, the localized learning rules, such as spike timing-dependent plasticity (STDP), simplify the data flow for learning and reduce the requirements for synaptic devices. However, unlike the mathematically proved powerful learning ability of the backpropagation algorithm in DNNs and their many recent applications, SNNs lack a developed learning algorithm that can be scaled up to be used in practical applications[29]. Deep belief networks (DBNs)[30] based on stacked restricted Boltzmann machines (RBMs) have binary activations and a localized learning rule, thus providing some of the additional advantages of SNNs while maintaining comparable performance in pattern recognition with backpropagation-powered DNNs.

In this Article, we report two-terminal floating-gate memristive synaptic devices — termed silicon synapses — that are fabricated in a standard CMOS production process flow. The silicon synapses exhibit self-selectivity, analogue conductance tunability, high retention time, small and predictable cycling degradation, and moderate device-to-device variation. The energy efficiency and performance density for multiply–accumulate (MAC) operations in an array is estimated to be 35.6 TOPs s$^{-1}$ W$^{-1}$ and 6.31 TOPs s$^{-1}$ mm$^{-2}$, which approximately 350 times and 170 times higher than a typical graphical processing unit (GPU). We use two 12-by-8 arrays of the memristive devices for in-situ training of a 19-by-8 memristive RBM for pattern recognition via a contrastive divergence based gradient descent algorithm. The neural states are fully binarized, thus simplifying the neural circuit design, with ADCs and DACs no longer needed. We then extend this approach to a memristive DBN that consists of three memristive RBMs for the training and recognition of the MNIST dataset. The layer-by-layer greedy learning of RBMs ensures that the weight updates only depend on local information. The performance of our system is equivalent to software approaches and comparable to mainstream memristive DNNs based on an error backpropagation algorithm, where complex neural behavior (the peripheral circuit of the memristive array) needs to be implemented in software or hardware.

**Two-terminal floating-gate memristive device**

Figures 1a and 1b show the schematic diagram and scanning electron microscopy (SEM) top view of the two-terminal floating-gate memristive device. The device consists of two parallelly connected transistors, i.e., the read transistor and the injection transistor, which share a common polysilicon floating gate (poly FG) and a drain (D). The sources of both transistors are connected externally (S). The device, i.e., the silicon synapse, is fabricated in a commercial 180nm CMOS process (Tower Semiconductor) without additional masks[31,32]. At positive bias ($V_D>0$ and $V_S=0$, thus $V_{DS}=V_D-V_S>0$), the device shows a non-linear current-voltage ($I_{DS}$-$V_{DS}$) relation as the combining result of the direct drain-source voltage and indirect floating gate voltage changes (Figure 1c). The device shows negligible current at reversed voltage bias ($V_D=0$ and $V_S>0$, thus $V_{DS}=V_D-V_S<0$) since the floating gate is coupled to the drain, and the transistors are turned off. The non-linear I-V behavior and low reverse current endow the device with a self-selection function and lead to low sneak-path currents in crossbar arrays, eliminating the need for the selector device[23] or operation of the array in half-voltage selection mode[33]. Different memory states can be obtained for different amounts of the charge stored in the floating gate, and thus different turn-on voltages of the transistors. In this work, we use the conductance $G$ of the device at 2V ($G=I_R/V_R$, $V_R$=2V) to denote the state of the memristive device. The device works in the subthreshold region, ensuring low power consumption for the reading operation.

The silicon synapse can be depressed (programmed) or potentiated (erased) by applying a voltage pulse on the drain or source terminal, which will inject electrons and holes into the floating gate[34–36], respectively (see Methods and Extended Data Fig. 1). Analog conductance tunability can be obtained by using depression or potentiation pulses with a width of 10 $\mu$s (Extended Data Fig. 2). The conductance states are nonvolatile (Extended Data Fig. 3). The depression and potentiation cycling test is performed by alternatively depressing and potentiating the device between the high conductance state (HCS, ~1$\mu$S) and the low conductance state (LCS, ~1nS). One loop of switching from the HCS to LCS and backward to HCS is defined as one test cycle. Up to 400 reliable test cycles have been measured with more than 53,000 writing pulses (Figure 1d), which shows small degradations as the later cycles need more pulses to switch between HCS and LCS (Extended Data Fig. 4). A physical-based model has been developed to simulate the switching behavior, including cycling degradation[37] (see Methods, Extended Data Fig. 1 and Extended Data Fig. 4). For digital switching (between HCS and LCS only), the endurance can be more than $10^5$ cycles[34].

To confirm the long-term retention in the analog switching, 16 devices are written to 16 evenly spaced conductance levels on the logarithmic scale between the LCS and HCS and monitored for more than two months at room temperature (Figure 1e). The LCS experiences small shifts, but none of the states are overlapping. The retention of analog conductance is projected to be higher than ten years, as indicated by the dashed lines in Figure 1e.

The two terminal floating-gate memristor has a radically different structure compared with conventional floating-gate transistors which have been proposed as synaptic devices for the biological learning following spike-timing-dependent plasticity (STDP) rule[38,39]. The simplified structure without a control gate simplifies the fabrication process and reduces the cell area, while the two-terminal structure greatly simplifies the potentiation, depression, and read operations[31].

**Memristive synaptic array.**

The silicon synapses are integrated into a crossbar array, accommodating 12-by-8 (thus 96 in total) memristors in a single chip (Figure 2a-2c). A test system, consisting of a customized PCB board, an FPGA evaluation board, and a B1500 semiconductor analyzer, is developed to enable individual memristors' reads, potentiations, and depressions in two 12-by-8 arrays (two chips, Figure 3a and 3b). The read operation of a single device can be performed by applying the read voltage ($V_D$=2V) on the D and grounding the S of the target device and floating all other terminals in the array. The low conductance of the device at reversed bias in Figure 1c provides the self-selection functionality and guarantees that there are no detectable sneak path currents.

The weighted summation of synapses in a neural network is physically embodied by the vector-matrix multiplications (VMM) using the silicon synapse array. The forward VMM is conducted in the array by applying the read voltage vector across the drains and sensing the current on the sources (Figures 3c and 3d). Due to the non-ohmic I-V behavior of the device, only binary input vectors are viable for the VMM operations, which meets the requirement of memristive RBM and DBN. The VMM operation in Figure 3c can be written as

$$I_{Sj} = \sum_i^m V_R v_i G_{ij}, \qquad (1)$$

where $G_{ij}$ is the conductance of the memristive device at the *i*th column and *j*th row of the *m*-by-*n* array, $v_i$ is the element of the binary input vector, $V_R = 2V$ is the read voltage, and $I_{Sj}$ is the element of the output vector. Due to the unipolar conductivity of the device, the implementation of the backward VMM is not straightforward. It can be conducted in the array by using the input vector to indicate whether we switch the source to the ground or float it, biasing all the drains to the read voltage, and sensing all the drain currents, as shown in Figure 3d. The backward VMM operation can be written as

$$I_{Di} = \sum_j^n V_R h_i G_{ij}, \qquad (2)$$

where $h_j$ is the element of the binary input vector and $I_{Dj}$ is the element of the output vector.

The conductance of each of the synaptic devices can be configured by potentiation and depression pulses (Figures 3e and 3f). The program/depression of the device in an array is straightforward since only the devices with a high program voltage on the drains and their sources grounded will be programmed. The devices with the drain or source floating will be prevented from the program operations, as shown in Figure 3e. The erase/potentiation of the device in an array needs special care since the erase pulse on the source will affect all devices in the row, no matter if the drains of the devices are grounded or floating. The devices that do not need to be erased in the row, however, can be deselected by applying a deselection voltage ($V_E'$=3.5V) on the drains, as shown in Figure 3f.

As illustrated in Figure 2d, we can see that there is no write or read disturbs over other devices in the array when operating a single device. The device-to-device variations are measured by fully depressing and potentiating all cells in the array between HCS and LCS (Figure 2e-2f and Extended Data Fig. 6). Pixelated letters and digits of "TECHNION" and "asic2lab" are written into the two chips (192 devices in total), as shown in Figure 2g, to verify the conductance tunability and memory function. All devices we have measured are working well, demonstrating 100% yield. For the multiply-accumulate (MAC) operations in an array, the computational efficiency is estimated to be $3.56\times10^4$ GOPs s$^{-1}$W$^{-1}$ energy and $6.31\times10^3$ GOPs s$^{-1}$mm$^{-2}$ in area, which are approximately 350 times and 170 times improvement on energy efficiency and performance density, respectively, compared with Tesla V100 GPU in 12 nm technology (see Methods)[27].

By far we have demonstrated the memristor devices with analog tunability, high endurance, long retention time, predictable cycling degradation, moderate device-to-device variations, high yield, and high energy efficiency for MAC operation. However, there are still two remaining non-ideal behaviors for synaptic application, namely the non-ohmic current-voltage (I-V) reading behavior and the non-linear conductance update for increasing potentiation or depression stimuli. These issues are also faced by other types of memristive synaptic devices, hindering the hardware implementation of DNN in simple and straightforward ways[40]. The non-ohmic issue prevents the analog multiplication using Ohm's law, which can be solved by using pulse number or pulse width modulated input signals to represent analog inputs[41,42]. The non-linear conductance update behavior refers to the fact that the change of conductance or weight per pulse depends on the current conductance or weight of the synapse synapses[43,44]. As a result, iterative write-and-verify operation[28,45] is needed to ensure the synaptic weight is correctly written to the target for each training batch. These solutions incur large hardware and latency overheads of the DNN in both inference and training stages. In short, our device is suitable to implement the DNN with existing optimization techniques proposed in recent years. However, here, we propose another type of neural network, i.e., DBN, that is more suitable to be implemented by the reported memristive device.

Unlike the mainstream DNN, we demonstrate the memristive DBN consisting of stacked memristive RBM layers. The inherent binary neuron states obviate the non-ohmic reading issues since only a fixed read voltage is needed. However, transferring the software version to a memristive synaptic device-based hardware system is not intuitive. In software implementation of DBN, the batch training method is used[30,46], resulting in excessive memory of intermediate neuron states of all samples in the batch. Additionally, the weight update in the software implementation is performed by adding or subtracting a small floating-point value on a floating-point number with high precision. However, high precision tuning of the conductance of the memristive devices is hard to be achieved. Ideally, we want to send a single program or erase pulse to depress or potentiate the conductance of the device (i.e., the weight), wishing that the same pulse will always result in the same amount of weight change. However, the non-linear weight update behavior exists. We also want to minimize the number of weight update operations, to save energy consumption and lose the requirement for the endurance of the device. Thus, we used a mixed-signal design and modified contrastive divergence learning algorithm for the training process[47]. The memristive RBM and DBN utilize a modified contrastive divergence learning algorithm that separates the synaptic weighted summation from the accumulation of weight update requests, simultaneously solving the non-linear weight update issue.

**Training of the memristive RBM**

We train a memristive RBM using two-terminal floating gate memristive devices as synaptic connections. Figure 4a shows the structure of the RBM, which has 19 visible units and 8 hidden units. The 19 visible units are partitioned into two parts: 12 for the pixel inputs of 4-by-3 pixelated letters or digits and 7 for the one-hot label inputs. Seven black-and-white patterns, each having 4-by-3 binary pixels ('1' for black pixels, and '0' for white pixels) denoting letters "A", "B", "C", "X", "Y", "0", and "1", are used as inputs (Figure 4b), and their labels are presented by one-hot vectors (for instance, the label for "A" has the code of "1000000"). The weight matrix between visible units and hidden units is thus partitioned into two sub-matrices: $w_1$ of the size 12-by-8 and $w_2$ of the size 7-by-8, which are implemented by two memristive device arrays ($G_1$ and $G_2$) in the two chips of our measurement system, as shown in Figure 4c.

The details of the training procedure of the memristive RBM are described in the Methods and Extended Data Fig. 7. The VMM and stochastic sampling are performed *in-situ* in the memristive device array with the help of externally injected noise currents[48] (Figure 4d). Intrinsic noise may also be exploited if the read noise is sufficiently large to support the stochastic sampling[48]. The inputs of the VMM and the stochastic sampling outputs are binary, thus no high-precision digital-to-analog or analog-to-digital converters are required. Forward VMM and sampling hidden units upon visible units, as well as backward VMM and sampling visible units upon the hidden units, are alternately performed, which gives the visible unit states $v$ (i.e., the input), the hidden unit states $h$, the reconstructed visible unit states $v'$, and the reconstructed hidden unit states $h'$. The contrastive divergence (CD) between the initial visible-hidden unit pairs and the reconstructed visible-hidden unit pairs, that is, $vh^T - v'h'^T$, provides the gradient descent for weight updates, that is, $\Delta w \sim (vh^T - v'h'^T)$, to minimize the energy of the RBM and improve the recognition accuracy in the inference stage.

The contrastive divergence has only ternary elements (-1, 0, or 1) since all the neuron unit states are binary. Here, we do not directly update the weight according to the instant contrastive divergence. Alternatively, the contrastive divergence elements are then accumulated in a digital counter array to avoid writing the device in every training sample input (Figure 4e). After the accumulated contrastive divergence element reaches a positive or negative threshold (+5 or -5 for the presented experiment), potentiation or depression pulses are sent to the memristive device to update the weights, respectively. The blind write method is used, which means that the updated weights do not need to be verified or fine-tuned, making the write operation to be fast and affordable. The memristive RBM is trained for 200 epochs, and each epoch comprises the 7 patterns and their labels, i.e., 7 training samples. Figures 4f and 4g show three representative traces of the accumulated contrastive divergence and conductance of corresponding memristive devices, respectively, as a function of the number of training samples. It can be seen that most of the consecutive weight updates cancel each other, leading to infrequent memristive device writing. The sparse write operations ensure low power consumption during neural network training. Figure 4h shows the conductance evolutions of all the devices in the two arrays during the training. The success of the training is reflected by the reduction of reconstruction error of the visible unit states ($<\boldsymbol{v}-\boldsymbol{v'}>$) (Figure 4i) and the increase of the recognition accuracy for all 7 learned patterns (Figure 4j), as a function of the training epoch. For a single training sample, the two forward and one backward VMM take 3 clock cycles. The weight updates on average take less than one cycle since less than one weight update operation is needed for each input sample. Thus, in total, less than four clock cycles are needed for processing one single training sample.

The neuron circuit has been the last missing piece that needs to be implemented in an integrated circuit due to the high complexity of the neural function. For instance, P. Yao et al. [28] implemented the neurons by using ADCs to convert the analog weighted output synaptic output to a digital signal, then using an ARM core to conduct the activation function, and finally using a DAC to pass the neuron output to synapses in the next layers. F. Kiani et al. [49] designed and implemented analog neuron circuits using on-the-shelf stand-alone components. In this work, however, the network structure and learning algorithm greatly simplify the neural circuit (Extended Data Fig. 8) which could be easily implemented in an integrated circuit.

**Training of the memristive DBN**

The single memristive RBM is extrapolated to memristive DBN consisting of three RBMs with the sizes of (784-500, 500-500, (500+10)-2000) stacking on each other (Figure 5a and Methods). The handwritten digits of the MNIST dataset[50] are used for training. Greedy learning algorithm[30] is employed here: we first train the RBM-1 layer using the digit images as input, followed by training the RBM-2 and RBM-3 layers sequentially using the hidden unit states of the previous layer and labels (last layer only) as input. The training of each layer follows a similar procedure as demonstrated in the previous section of the single RBM layer. For the first layer, the 784 input nodes receive input for each pixel of the digit images with the size of 28x28. The pixels of the grayscale images are firstly normalized to the range of [0, 1]. Then, the normalized values were taken as the probability of sampling pixels to be 1s or 0s, constituting the binary input vector. The reconstruction errors for each RBM layer are shown in Figure 5b, revealing apparent reduction and convergence of the reconstruction error. The image recognition, i.e., the inference of the neural network, can be conducted by expanding the RBM-3 layer, i.e., processing the image by the path $w_1 \rightarrow w_2 \rightarrow w_3 \rightarrow w_4$. Since each RBM layer is designed to have a stochastic binarized output for the training, the test accuracy has a probabilistic distribution. The stochasticity can be eliminated by removing the noise current injection to speed up testing in Figure 4d, such that the inference is *deterministic*. Another way is to repeat the stochastic inference and take the average (*sampling inference*). Figure 5c shows the test accuracy of the test set images in the MNIST dataset during the training of the memristive DBN for both deterministic inference and 50-time sampling inference. The first 30 epochs are in the phase of the training of the RBM-3, and the last 30 epochs are in the phase of fine-tuning the memristive DBN by the wake-sleep algorithm[51] under the same hardware design (see Methods). The accuracy of deterministic inference after the 60-epoch training is 95.67%, while the 50-time sampling inference gives an accuracy of 97.05%, which are 2.63% and 1.25% lower than the DBN trained in software (98.3%), respectively. The accuracy for the 50-time sampling inference (Samp. #50) is comparable to recently reported DNN based on PCM[27,52] and RRAM[28] devices where the peripheral circuit to realize the neural activities are much more complex to be realized in hardware, as summarized in Figure 5d.

**Conclusion**

We have reported CMOS-compatible silicon synapses that are based on two-terminal floating-gate memristive devices. The synapses offer self-selection, analogue conductance tunability, reliable cycling behavior, high yield, moderate device-to-device variation, predictable cycle to cycle degradation, and low energy consumption. We used the silicon synapses to train a memristive RBM. The memristive RBM features binary neurons, thus removing non-Ohmic I-V related issues and reducing the complexity of the peripheral (neural) circuitry. Using a contrastive divergence based gradient descent learning algorithm, the calculation and accumulation of the weight update requests are separated from VMM and sampling operations in the RBM in a mixed-signal learning scheme, making the training immune to the non-linear weight update behavior. As a result, the number of write operations, and thus the endurance requirement of the memristor device, is greatly reduced. We also extended the memristive RBM to create a DBN, which exhibits a performance that is comparable to software and state-of-the-art memristive DNNs that require more complex neural circuits.

**METHODS**

**Device fabrication and characterization.** The memristive devices were fabricated in Tower Semiconductor's 180nm/5V CMOS process (gate oxide thickness 110 Å). Each device consists of two NMOS transistors: the readout transistor and the injection transistor (Figure 1a-1b, and Extended Data Fig. 1a), sharing a common floating gate using the polysilicon layer – the normal gate layer for CMOS transistors. The floating gate is surrounded by high-quality dielectrics, ensuring high retention of the charge stored in the floating gate. No additional masks are needed for the fabrication compared with the standard CMOS process. The injection transistor has a shorter channel (0.3 $\mu$m) and an additional p-type implant in the drain junction (compared with standard NMOS transistors), optimized to enhance the hot carrier injection. The read transistor (channel length 0.6$\mu$m) has a lower threshold voltage to enhance the read current. The device structure was originally developed as an embedded digital nonvolatile memory (NVM), named as Y-Flash memory[32], and is widely used by Tower Semiconductor foundry customers.

For single device measurement, the DC sweeping measurement and pulse-induced switching characteristics of the device were conducted by probing the device using Cascade 12000 probe station and collecting the data using a B1500 Semiconductor Analyzer. The source measurement unit (SMU) module of B1500 is used to conduct the DC sweeping measurement, and the waveform generator/fast measurement unit (WGFMU) is used to provide the pulse with accurate pulse width and amplitude for potentiation and depression operations. The substrate of the device is consistently grounded. The device should always be operated with positive voltages since a negative voltage on the drain or source terminals of the transistors will induce an unexpected large current from the ground due to the positive bias of the p-n junctions between the p-type substrate and n-type ohmic contacts of the drain/source terminals (Extended Data Fig. 1a). The sources of the readout transistor (Source Read, SR) and the injection transistor (Source Injection, SI) are shortened externally (Extended Data Fig. 1a). Thus, the device acts as a two-terminal memristive device (Extended Data Fig. 1b).

**Readout, depression, and potentiation.** The readout operation is conducted by applying a voltage pulse (amplitude $V_R$=2V) and sensing the current ($I_R$), as shown in Extended Data Fig. 2a. The state of the device is denoted by the conductance $G=I_R/V_R$.

The depression of the memristive device corresponds to the program operation of the conventional floating gate device. It is conducted by applying a high voltage pulse on the drain terminal and grounding the source terminal (Extended Data Fig. 2b). We used a fixed amplitude of $V_P$=5V for the depression/program operations. The potentiation of the memristive device corresponds to erase operation. It is conducted by applying a high voltage pulse on the source terminal and grounding or floating the drain terminal (Extended Data Fig. 2c). The fixed amplitude of $V_E$=8V is used for the potentiation/erase operations. We note that the Y-Flash memory shares the same write mechanism as common floating gate transistor devices, the high voltage operations are commonly required for the memory cells based on floating gate transistor devices[54]. However, for ASIC implementation, there are mature technologies to generate high voltages inside of the integrated chips, like charge pumps[54].

The pulse-induced depression is performed by alternatively applying the program pulse and readout pulse, as shown in Extended Data Fig. 2d. When program pulses have a short pulse width (10 us), we can obtain more than 1500 conductance states between HCS and LCS (Extended Data Fig. 2e), showing an analog type switching behavior. The pulse-induced potentiation is performed by alternatively applying the erase pulse and readout pulse, as shown in Extended Data Fig. 2d. Similar to the depression, when the erase pulses have a short pulse width (10 us, Extended Data Fig. 2f), we can obtain analog-type behavior with more than 800 conductance states from LCS to HCS (Extended Data Fig. 2g).

Varying the pulse width for the depression and potentiation operations trades off between a higher number of conductance states and a larger conductance change stride. We chose the pulse width for the depression and potentiation to be 200 us and 100 us to character the device and for tuning the conductance for neural network learning. The typical pulse-induced depression and potentiation curves are shown in Extended Data Fig. 3a and Extended Data Fig. 3b, respectively. We continuously read the device for all intermediate states for 20 seconds, and all the states are stable and show negligible read noise (Extended Data Fig. 3c and Extended Data Fig. 3d).

**Operation mechanisms and device modeling.** To model the I-V characteristics of the device, we first need to obtain the floating gate voltage ($V_{FG}$). The coupling capacitor net determines the floating gate voltage between the device's floating gated and other terminals and the charge $Q_{FG}$ at the floating gate. The floating gate voltage can be written as:

$$V_{FG} = \frac{Q_{FG}+C_{GD}V_D+(C_{GSR}+C_{GSI})V_S}{C_{GSR}+C_{GD}+C_{GSI}+C_{GB}} \approx \frac{Q_{FG}+C_{GD}V_D}{C_{GSR}+C_{GD}+C_{GSI}+C_{GB}}, \tag{3}$$

where $C_{GSR}$, $C_{GD}$, $C_{GSI}$, and $C_{GB}$ are the capacitance between the floating gate and source read, the floating gate and drain, the floating gate and source injection, and the floating gate and substrate, respectively, as shown in Extended Data Fig. 1a. $V_D$ and $V_S$ are the voltages applied on the drain and source terminals, respectively. Notice that the capacitance between the floating gate and the drain ($C_{GD}$) is much higher than other capacitances[26]. For positive bias of the device ($V_D$>0, $V_S = 0$), $V_{FG}$ linearly increases with $V_D$ and the transistors will be gradually opened (Extended Data Fig. 1c). The current of the device can be obtained by determining the drain-to-source current of both the readout transistor and injection transistor in both the sub-threshold and above-threshold regions. For reversed bias of the device ($V_D=0$, $V_S > 0$, thus $V_{DS}=V_D - V_S < 0$ ), $V_{FG}$ is defined primarily by drain voltage. The transistors will remain closed, and the device shows a low readout current.

A high drain-to-source current is expected when a higher voltage is applied to the drain and the source is grounded. In the injection transistor with a short channel, the high source injection-to-drain electrical field accelerates the electrons. A small number of accelerated electrons (hot electrons) will be injected into the floating gate, i.e., channel hot electron injection (CHEI), as shown in Extended Data Fig. 1d. The process is similar to the program operation in conventional floating gate devices[35,55]. The induced gate current can be written as

$$I_{G,program} = -I_{DS}P_0 e^{-\frac{V_\alpha}{V_{FG}}}, \qquad (4)$$

where $I_{DS}$ is the drain-to-source current of the injection transistor, $P_0$ is the probability of the hot electrons to be emitted into the floating gate, and $V_\alpha$ is a fitting parameter. The electron injections to the floating gate will decrease the total charge on the floating gate. Thus, the floating gate voltage is decreased for a particular read voltage bias. The readout current and the conductance of the device are decreased, corresponding to the depression of the synaptic weight.

The drain-source current is negligible when a higher voltage is applied to the sources, and the drain is grounded or floating. However, the barrier between the channel and the source in the injection transistor is reversely biased, which will induce the band-to-band hole generation in the drain region of the transistor. The high source-to-drain electrical field accelerates the generated holes, and the lucky ones are injected into the floating gate (Extended Data Fig. 1e). The hot hole gate current can be written as[56,57]

$$I_{G,erase} = \xi(V_{FG} - V_{bi})^2 e^{-\frac{\beta}{V_{FG}-V_{bi}}}, \qquad (5)$$

where $\xi$ and $\beta$ are the fitting parameters reflecting the hole injection efficiency, and $V_{bi}$ is the channel potential in the injection site, where the injection of holes takes place. This process decreases the total charge on the floating gate and increases the conductance of the device, corresponding to the potentiation of the synaptic weight.

The cycling degradation of the device is modeled by adding the degradation features to the parameters controlling the efficiency of electron or hole injections in the program and erase operations, i.e., $V_\alpha$ and $\beta$, respectively. The degradations of these two parameters are written as

$$\frac{dV_\alpha}{dt_{prog}} = (V_{\alpha,max} - V_\alpha)(e^{\frac{t_{prog}}{\tau_{prog}}} - 1), \qquad (6)$$

and

$$\frac{d\beta}{dt_{erase}} = (\beta_{max} - \beta)(e^{\frac{t_{erase}}{\tau_{erase}}} - 1), \qquad (7)$$

where $t_{prog}$ and $t_{erase}$ are the total time of the program or erase operations applied to the device,

$V_{\alpha,max}$ and $\beta_{max}$ are the stable values of these parameters after operating for a long time, $\tau_{prog}$ and $\tau_{erase}$ are the time constants of the degradation.

The increase of the full depression/program time and full potentiation/erase times as the function of the number of the cycles is shown in Extended Data Fig. 4, which shows good agreement between the data and the model. The simulated depression and potentiation curves match the experimental data well, as shown in Fig. 1d.

The device-to-device variation is modeled by adding a Gaussian-type variation to the parameters in the exponents of the gate current for the program and erase operations, that is, $V_{\alpha,d2d} \in \mathcal{N}(V_\alpha, \sigma_{V_\alpha}^2)$ and $\beta_{d2d} \in \mathcal{N}(\beta, \sigma_\beta^2)$, where $\sigma_{V_\alpha}$ and $\sigma_\beta$ are the standard variations of the parameters $V_\alpha$ and $\beta$. The values of the parameter $V_\alpha$ and $\beta$ for each device are randomly chosen from the distributions at the initial stage. The simulated program and erase traces, including device-to-device variations, are shown in Extended Data Fig. 6, which well reproduces the measure of device-to-device variation in Fig. 2e and Fig. 2f.

**Testing system.** To enable the investigation of device operation within an array, device-to-device variations, and demonstration of the training and inference of a memristive RBM layer, a test system which integrates two Y-Flash chips, each containing a 12-by-8 memristive device array, is developed and presented in Fig. 3a, Fig. 3b, and Extended Data Fig. 5. The system consists of a custom printed circuit board (PCB, a Z702 FPGA development board, a B1500 semiconductor analyzer, a controlling personal computer (PC), and power supplies.

In the custom PCB board, we use cross-point switch array (AD75019) chips to route all the terminals of the two Y-Flash chips to four channels of the B1500 SMUs/WGFMUs. The FPGA configures the cross-point switching array chips such that SMUs/WGFMUs can be arbitrarily connected to each terminal of the memristive device array. The WGFMUs provide the pulses for readout, potentiation, and depression operations and sense the current in the readout mode. The SMUs in B1500 provide voltage biases for different operation modes. The FPGA board and the B1500 are controlled by a PC to synchronize the connections and the pulse applying/sensing operations.

**Memristive RBM demonstration.** The testing system containing two 12-by-8 memristive device arrays is used to demonstrate the training and inference of the RBM in Figure 4a. The RBM has two weight matrices: $w_1$ with the size of 12-by-8 and $w_2$ with the size of 7-by-8, which are instantiated by the two memristive arrays, Chip 1 and Chip 2, respectively, as shown in Figure 4c. To represent the signed weights, we use a separated row and a separated column of the memristive devices with the conductance of $G_{ref}$ to subtract from the always positive conductance of the memristive array. For instance, as shown in Figure 4d about the forward VMM and stochastic sampling, the weight of each element in the weight matrix can be expressed as $w_{ij} = G_{ij} - G_{ref}$.

The training procedure of the memristive RBM is shown in Extended Data Fig. 7. The testing system performs VMM and weight updates in the memristive array. The stochastic sampling and the calculation and accumulation of the contrastive divergence are performed in software. For instance, the probability of the hidden units being excited is determined by

$$P(\mathbf{h} = 1) = \sigma(\mathbf{I_h}/I_0) \tag{8}$$

where $I_{hj} = \sum_i^m V_R v_i G_{ij} - \sum_i^m V_R v_i G_{ref}$ is the current differences of the output from the VMM operation on the memristive device array, $\sigma(\cdot)$ is the logistic function, and $I_0$ ($I_0 = 0.2\mu A$ in the RBM training) is the step parameter of the logistic function, which can be realized by controlling the intensity of the noise current[48] in Figure 4d. Note that although a PC is involved in the neural network operations, similar as most approaches in the demonstrations of memristive neuromorphic computing systems[11,58,59], the software operations by the PC can be fully implemented in hardware when integrated into an application-specific integrated circuit (ASIC).

**Memristive DBN simulation.** The simulation of the learning and inference of the memristive DBN uses the model of the memristive device, which nicely captures the device I-V behaviors, potentiation and depression conductance updating behavior, cycling degradation, and device-to-device variations. The contrastive divergence threshold for the training is set to 64 and -64, corresponding to signed 6-bit digital counters in the contrastive divergence accumulation array. The steep parameter of the logistic function in Eq. 8 is set to 1 uA ($I_0 = 1\mu A$).

**DBN fine-tuning.** A fine-tuning procedure of the memristive DBN is performed using the wake-sleep algorithm[30,51] after the greedy-learning of the DBN (pretraining). To do the fine-tuning, the RBMs in the memristive DBN are duplicated, except for the last RBM, with the new weight matrices $w_1 = w_1'$ and $w_2 = w_2'$ (Figure 5a). The weight matrices $w_1$, $w_2$, $w_3$, and $w_4$ constitute the recognition path and the weight matrices $w_4$, $w_3$, $w_2'$, and $w_1'$ constitute the generation path or wake path. The input data of the image and the label are fed to the recognition path, and the top RBM layer performs Gibb sampling iteratively (iteration number is 20 in our simulation), and then the reconstructed visible neuron states are fed to the generation path to generate the reconstructed image (sleep path). The states of neurons in the wake path and sleep path are used to update the weight matrix in the sleep path and wake path, respectively. For the simulation results shown in Figure 5, the DBN is pre-trained for 30 epochs and fine-tuned for another 30 epochs.

**Design of peripheral circuits.** The designed memristive RBM and DBN utilize mixed-signal learning: the memristive array conducts the VMMs and the stochastic sampling in the analog domain, and other parts of the system, including the contrastive divergence accumulation and communication between adjacent RBM layers, work in the digital domain. The full peripheral circuit of the memristive array in the analog domain is designed and shown in Extended Data Fig. 8. The peripheral circuit includes multiplexers, trans-impedance amplifiers, noise current generators, and comparators. No digital-to-analog converters (DACs) or analog-to-digital converters (ADCs) are needed in the peripheral circuit.

**Metric estimations.** A single device in an array has a footprint of 2.7 $\mu$m ×1.17 $\mu$m = 3.159 $\mu m^2$ (180 nm technology). The maximum energy consumption of the readout, program, and erase operations are estimated as 2V×1uA×50ns = 0.1 pJ, 5V×20uA×200us = 20 nJ, and 8V×10nA×100us = 8 pJ, respectively. Note that the actual energy consumption should be calculated according to the actual conductance states of the devices, which can be obtained from the neural network simulation. The weight update operations consume much more energy than the read operations since they need longer pulses, and the program/depression operations work in the transistor-on mode. However, since we use the contrastive divergence array to accumulate the required weight updates, the number of weight update operations is substantially reduced, thus reducing the energy consumption. In the training of the memristive DBN shown in Figure 4, the VMM operations consume 18 nJ per training image, while the weight updates consume 33 nJ per training image for the first RBM layer (size: 784-by-500). This means that approximately only three weight update operations (33nJ/(20nJ + 8pJ)×2) are performed for each image input. The energy consumption of the other layers is proportional to their array sizes.

A single VMM operation on the first RBM layer consumes 11 nJ and takes 50 ns, which performs 784×500=392,000 multiply–accumulate (MAC) operations. This results in energy efficiency of $3.56\times10^4$ GOPs $s^{-1}W^{-1}$. The area of the memristive array is 3.159 $\mu m^2$×(784+1)×(500+1)≈1.24 $mm^2$, which gives the performance density of $6.31\times10^3$ GOPs $s^{-1}mm^{-2}$. These metrics show roughly 350 times and 170 times improvement in energy efficiency and performance density, respectively, compared with Tesla V100 GPU in 12 nm technology (energy efficiency: 100 GOP $s^{-1}W^{-1}$, performance density: 37 GOP $s^{-1}mm^{-2}$ for 16-bit floating-point number computing)[27]. Note that the energy consumption and footprint of the peripheral circuits are not included in the calculation. Also, note that scaling down the device will strongly improve the metrics.

**DATA AVAILABILITY**

The data that support the plots within this paper and other finding of this study are available as Source Data files associated with the Figures and Extended Data Figures.

## CODE AVAILABILITY

The code that supports the device modeling and neural network simulations in this study is provided as Supplementary Source Code and is also available at [https://github.com/wangweifcc/memristive_dbn_yflash](https://github.com/wangweifcc/memristive_dbn_yflash).

## ACKNOWLEDGMENTS

This work was supported by the European Research Council through the European Union's Horizon 2020 Research and Innovation Programe under Grant 757259 and FET-Open NeuChip project under grant agreement No. 964877. W.W. was supported in part at the Technion by the Aly Kaufman Fellowship. W.W. acknowledge the help from Devangshu Dutta on the PCB and FPGA code development.

## AUTHOR CONTRIBUTIONS

W.W. conceived the concept of Y-Flash memristor-based RBM and DBN, L.D. designed the memristive chip and led tapeout to fabrication, including array layout and readout cells, and developed device level operation schemes, Y.R and E.P. suggested the Y-Flash memristor cell and performed initial verification, W.W. conducted the device characterization, array level operation schemes with the assistance of L.D., E.H., and B.H. W.W. conducted neural network demonstration, and simulations with the assistance of L.D. and Y.L. Y.R., E.P., and Z.W. helped the illustration results. All of the authors discussed the results and contributed to the preparation of the manuscript. S.K. supervised the research.

**Competing interests**: The authors declare no competing interests.

# FIGURES AND FIGURE CAPTIONS

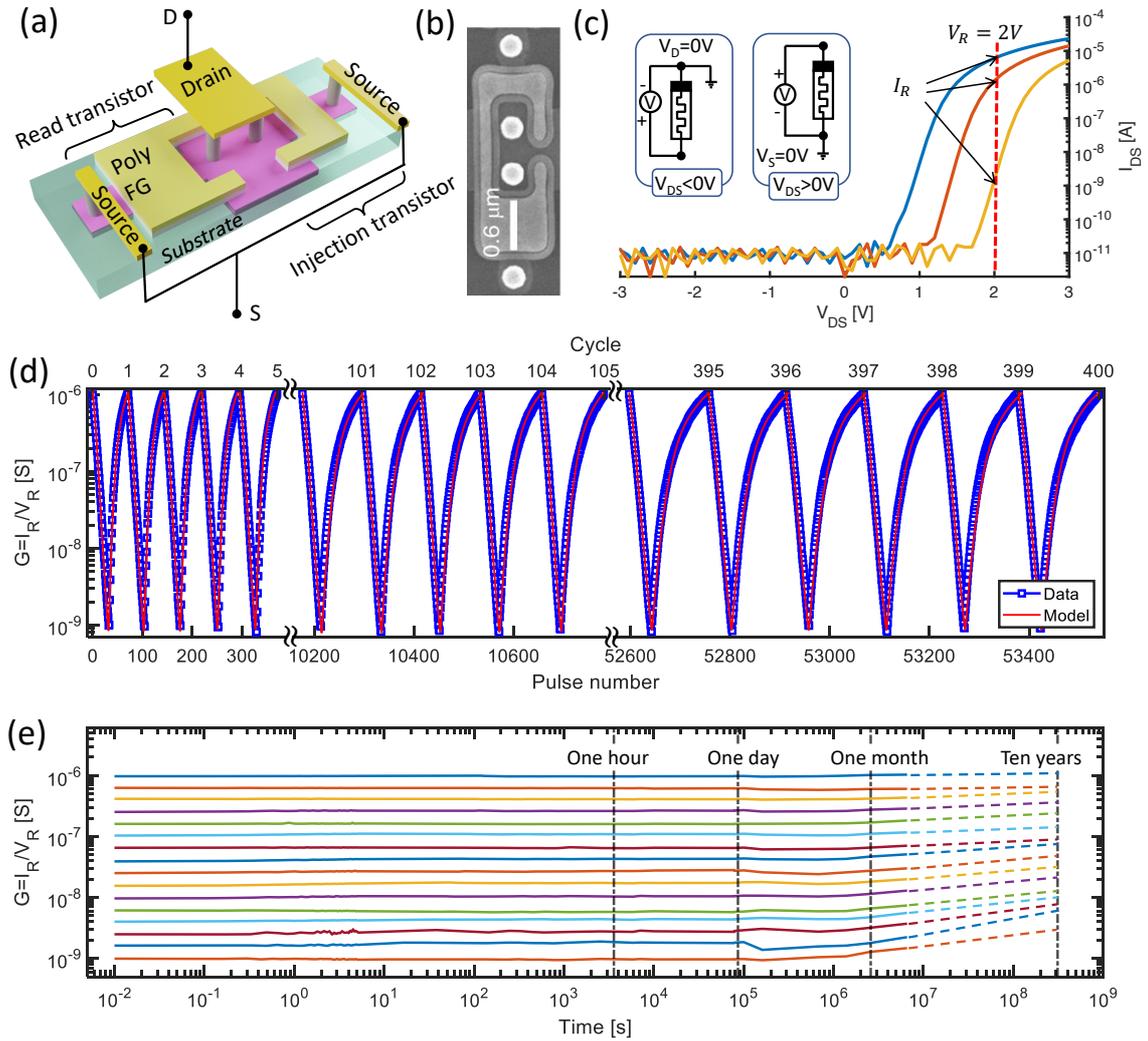

**Figure 1 | Memristive synapse based on two-terminal floating gate devices.** (**a**) Illustration of the floating gate devices featuring parallel connection of two transistors with a common floating gate and a common drain. The two sources are connected to form a two-terminal configuration. (**b**) Scanning electron microscopy (SEM) image of the fabricated device. (**c**) The low voltage sweep reading of the devices shows high self-selection ability. The states (different colored lines) of the device are defined by the current ($I_R$) at positive bias ($V_R=2V$). Inset: schematics of the negative and positive bias reading configurations (the I-V curves are measured in the two configurations separately and concatenated in the main graph). (**d**) Write cycling of the device. The device is alternatively depressed and potentiated between the high conductance state (HCS) and the low conductance state (LCS) for 400 test cycles (see Methods, Extended Data Fig. 2, and Extended Data Fig. 3 for the details of the write operations). A compact model is developed to accurately describe the conductance switching and cycling degradation (Methods and Extended Data Fig. 4). (**e**) Retention of 16 conductance levels of the devices measured for more than one month at room temperature and projected to 10 years.

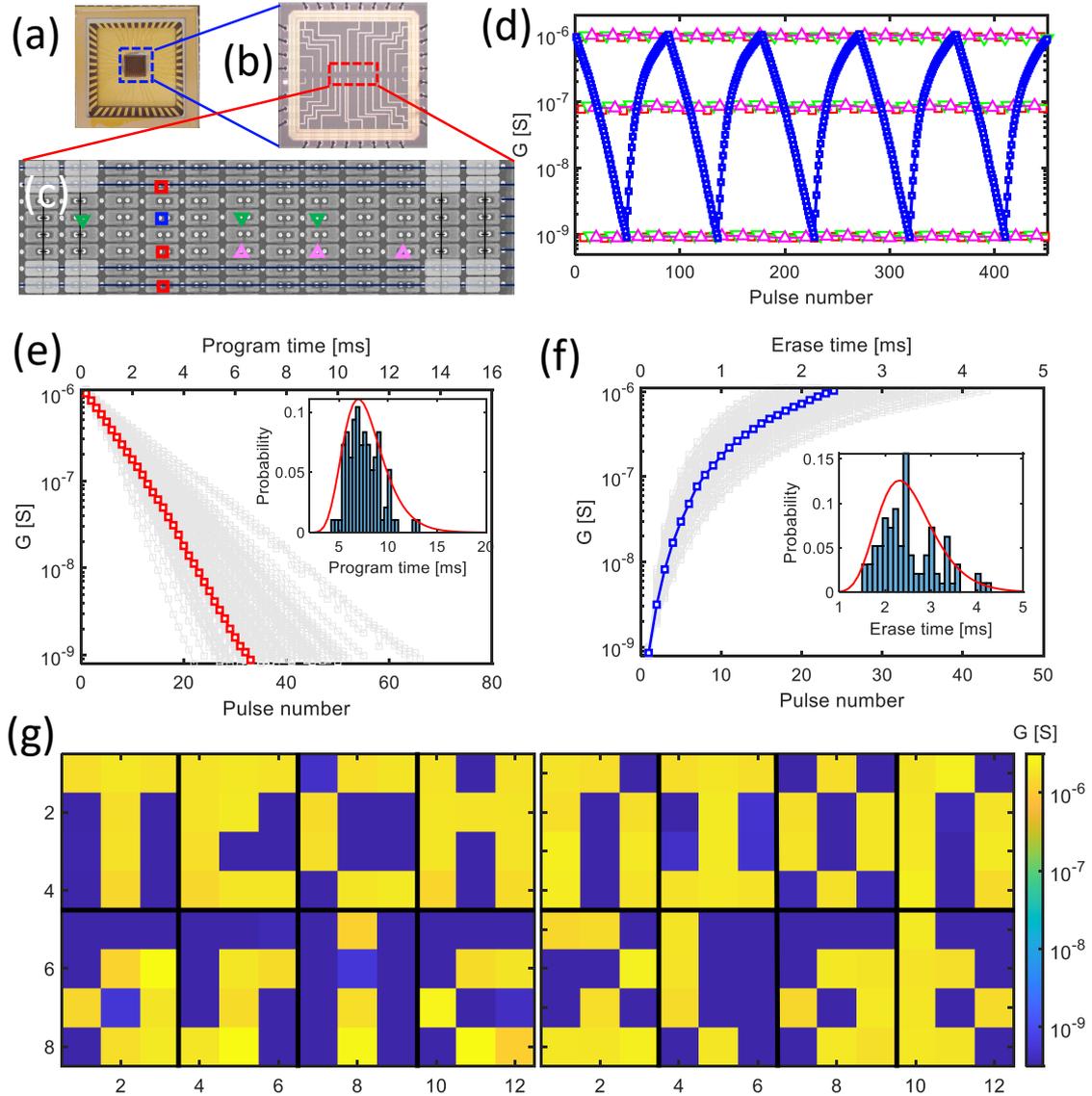

**Figure 2 | Memristive synaptic array and device-to-device variations.** Images of (**a**) the packaged chip and a (**b**) the 12x8 size array design. (**c**) SEM image of the 12x8 array. (**d**) Write disturb of silicon synapses in the same array when writing to a single selected device (device locations are indicated in **c**). (**e**) Depression behavior of all the 96 devices in an array. (**f**) Potentiation behavior of all the 96 devices in an array. Insets of (e) and (f): statistical results of the full program time from HCS to LCS and the full erase time from LCS to HCS, respectively. The model well captures the device-to-device variation (see Methods and Extended Data Fig. 6). (**g**) Demonstration of the memory function by writing patterns of letters and digits ("TECHNION" and "asic2lab") into two 12x8 memristive arrays.

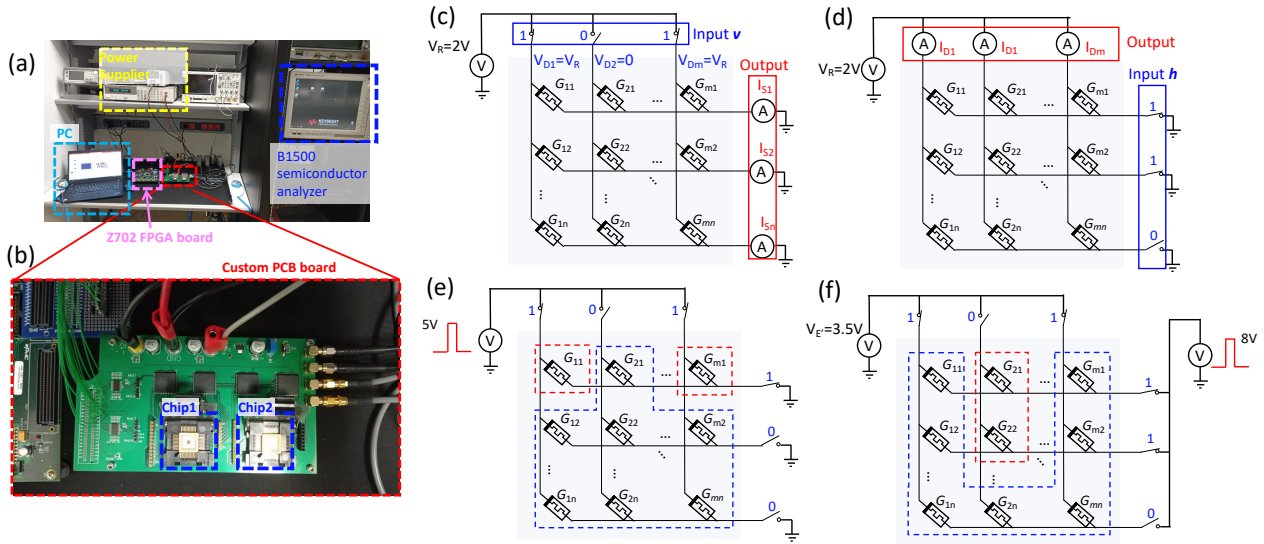

**Figure 3 | The setup of the test system for array operations and memristive RBM demonstration.** (a) Overview of the system consisting of a custom printed circuit board (PCB) board, a Z702 FPGA development board, a B1500 semiconductor analyzer, a controlling personal computer (PC), and power suppliers. (b) The details of the custom PCB board, which integrates two Y-Flash chips, each of which has a 12-by-8 array of two-terminal floating gate memristive devices. Schematics of performing (c) the forward VMM and (d) the backward VMM operations on an array of the two-terminal floating gate memristive devices. Schematics of performing (e) the depression/program operation and (f) the potentiation/erase operations on selected devices (marked by the red dashed line) in an array of the memristive devices.

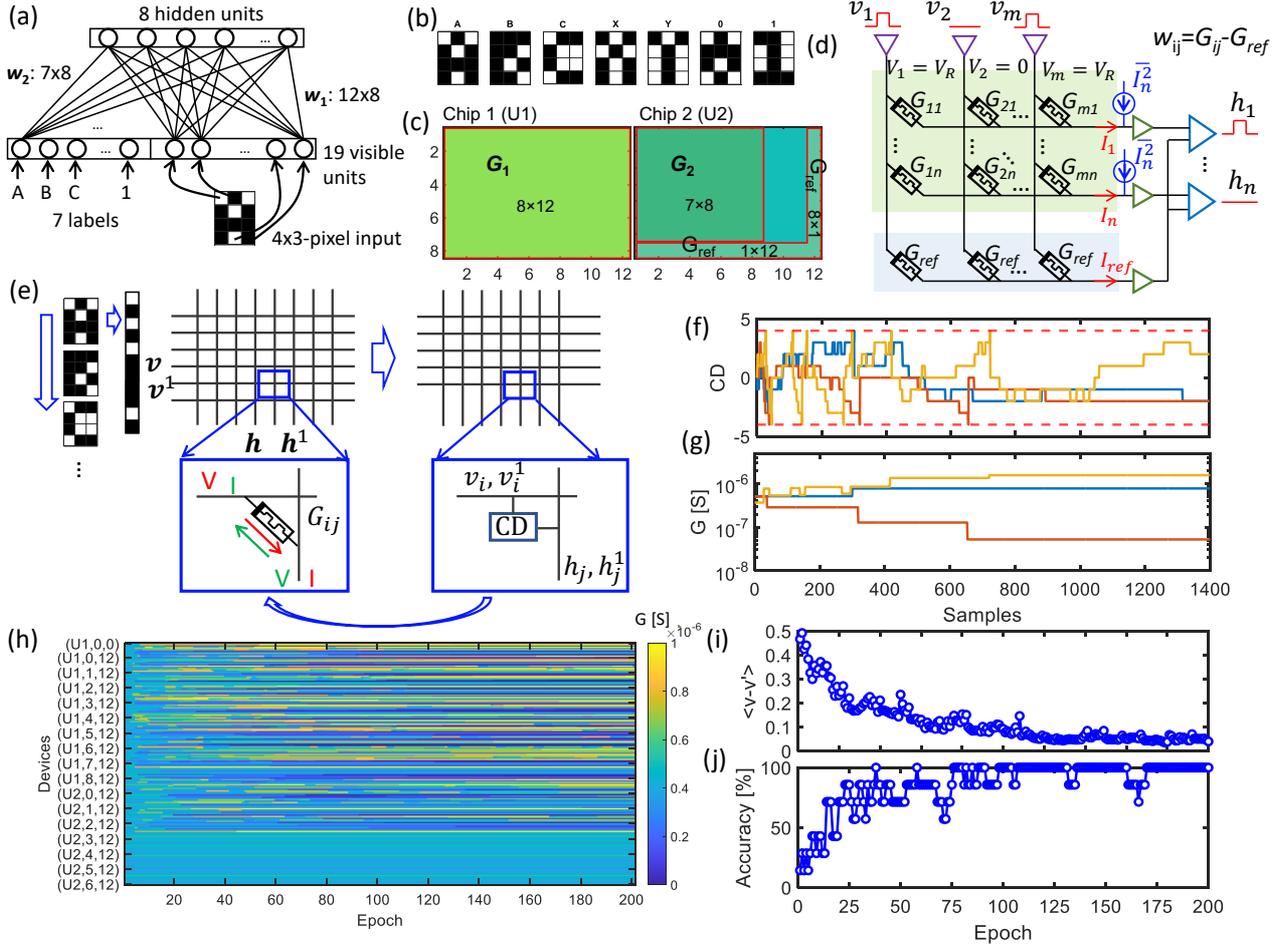

**Figure 4 | Demonstration of the restricted Boltzmann machine (RBM) training using two memristive synaptic chips**. (**a**) The RBM structure for demonstration with 19 visible units (12 for the pixels of input images and 7 for labels) and 8 hidden units. (**b**) Images of 7 image patterns of 4x3 size for the training. (**c**) Allocation of the weight matrix on the two memristive synaptic chips. (**d**) Schematic illustration of performing vector-matrix multiplications and stochastic sampling (sampling VMM) within the synaptic arrays. (**e**) Schematic of the mixed-signal training of an RBM by separating the sampling VMM and gradient accumulation. Detailed evolutions of (**f**) the accumulation of contrastive divergence (CD) and (**g**) the conductance of the synaptic device during the training. (**h**) Conductance evolution of all synaptic devices during the training. (**i**) The reconstruction error of RBM as a function of the training epoch. (**j**) Test accuracy as a function of the training epoch.

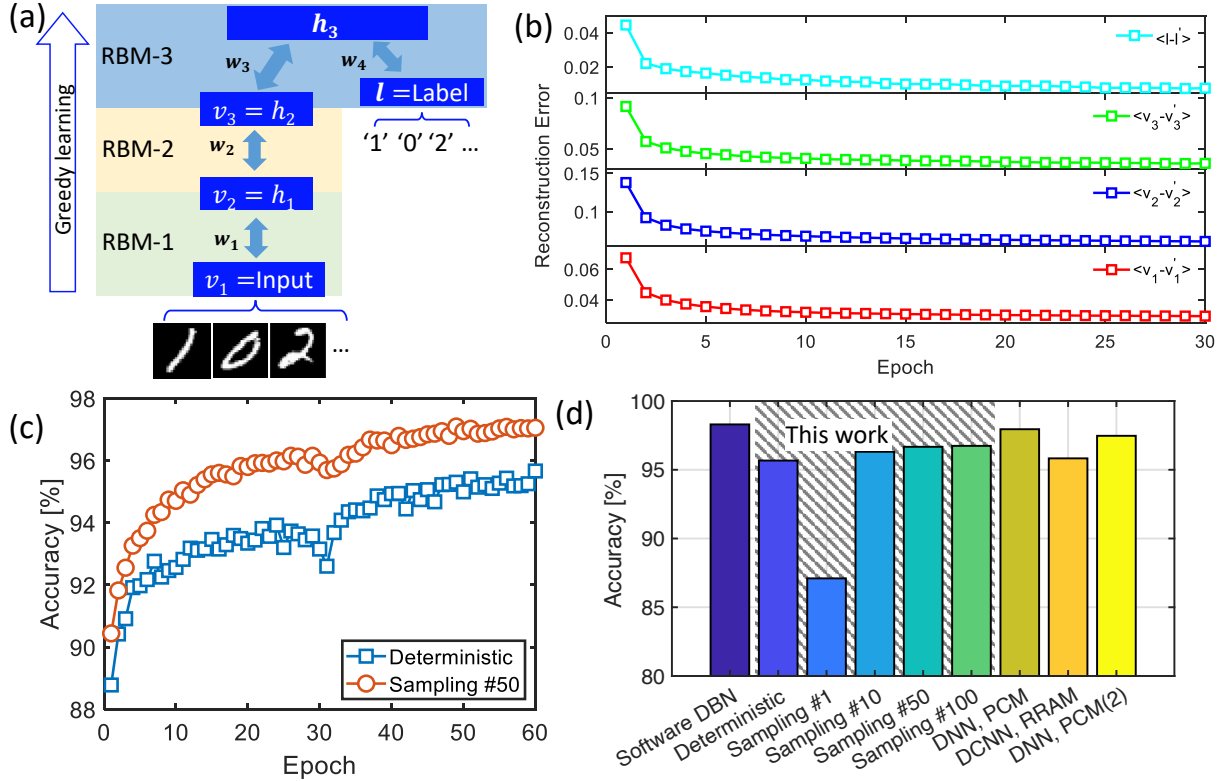

**Figure 5 | Training of the deep belief net based on the silicon synapses for MNIST dataset**. (**a**) Structure of the deep belief net consisting of stacked RBMs. (**b**) The reconstruction errors of the visible units in the three RBMs as a function of the training epoch. (**c**) Test accuracy of the neural network as a function of training epoch for deterministic inference and sampling inference of 50 times repetition. (**d**) Comparison of the test accuracy of the floating gate memristive device-based deep belief net with the software solution and other memristive deep neural networks [DNN-PCM[27], DCNN-RRAM[28], DNN-PCM(2)[53]].

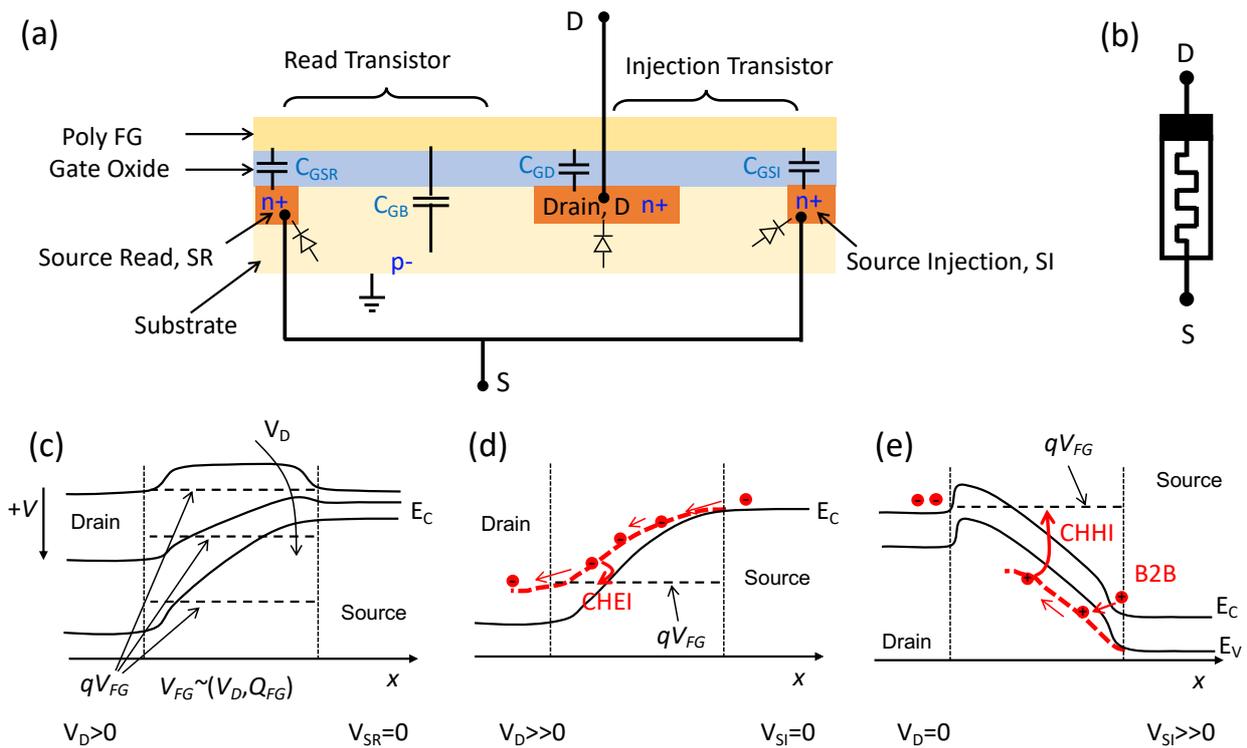

**Extended Data Fig. 1 | Device structure and band diagrams in different operational modes.** (a) Schematic illustration of the device structure from the sectional view. The injection transistor (shorter channel and higher threshold voltage) and the read transistor (longer channel and lower threshold voltage) are parallelly connected: they have a common drain and a common floating gate; their sources are externally shortened. The capacitance between the floating gate and the drain ($C_{GD}$ is much larger than other capacitances), thus the floating gate voltage is mainly coupled to the drain. (b) The equivalent circuit symbol of the device. (c) Band diagrams of the read transistor in the reading mode. A low drain voltage will induce a low floating gate voltage and keep the transistor closed. As the drain voltage increases, the floating gate voltage increases, and the transistor gradually opens. The injection transistor shares similar behavior but has a lower current since the threshold voltage is higher. (d) The band diagram of the injection transistor at the depression (program) mode. When a higher voltage is applied to the drain and the source is grounded, the electrons in the channel are accelerated by the electrical field in the drain region, and the lucky ones are injected into the floating gate, i.e., channel hot electron injection (CHEI). (e) The band diagram of the injection transistor in the potentiation (erase) mode. When a higher voltage is applied to the source and the drain is grounded or floating, the source p-n junction is reversely biased. This induces band-to-band (B2B) hole generation in the source region. The high lateral electric field accelerates the generated holes, and the lucky ones are injected into the floating gate. The hot electron/hole effects are negligible in the readout transistor.

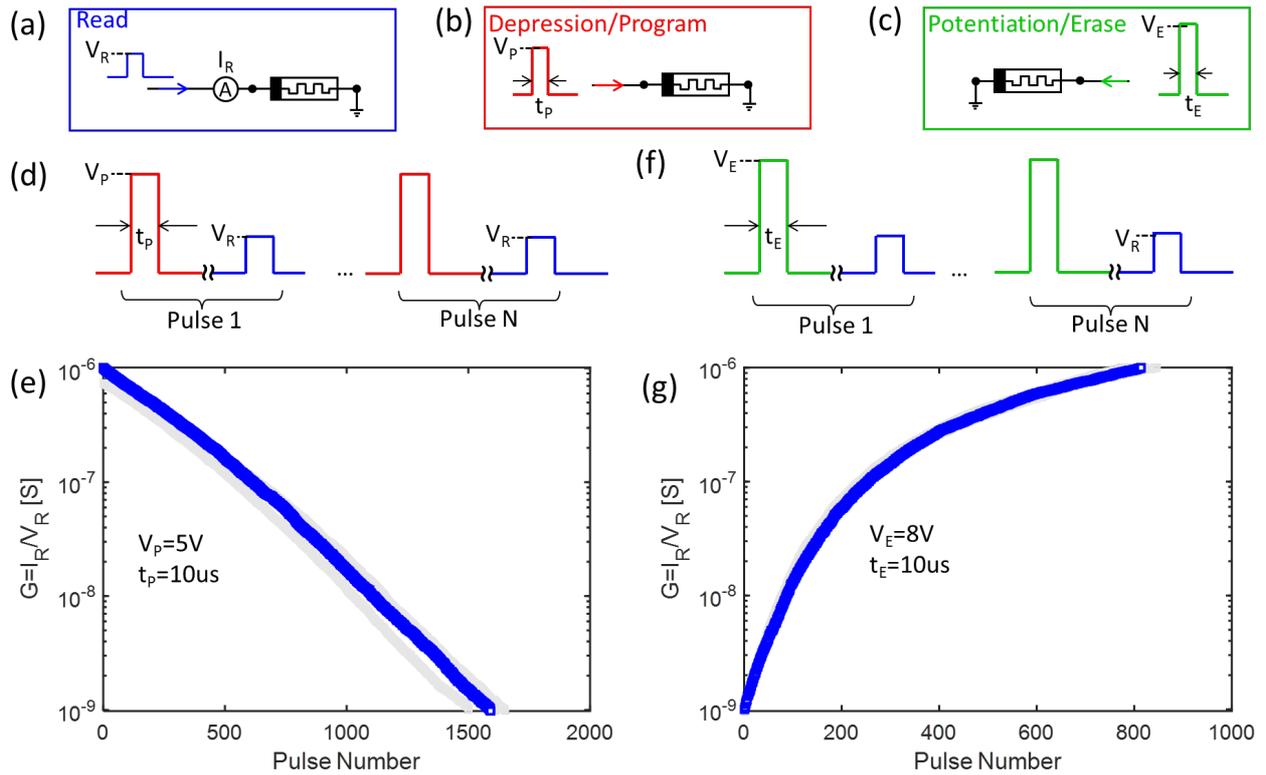

**Extended Data Fig. 2 | Depression and potentiation of the devices by continuous program or erase pulses.** (a) Read mode of the device by applying a voltage pulse ($V_R$=2V for pulse reading) on the D terminal and sensing the current ($I_R$). (b) Depression (program) mode of the device operation by applying a voltage pulse ($V_P$=5V) on the D terminal and grounding the S terminal. (c) Potentiation (erase) mode of the device operation by applying a voltage pulse ($V_E$=8V) on the S terminal and grounding, or floating, the D terminal. The floating D terminal will be capacitively coupled to the grounded substrate. (d) Schematic of the pulse depression/program test by alternatively applying the program pulse and reading the device by the read pulse. (e) The device conductance ($G=I_R/V_R$) as a function of the pulse number when depressing the device by programming pulses with the width of 10 us. (f) Schematic of the pulse potentiation/erase test by alternatively applying the erase pulse and reading the device by the read pulse. (g) The device conductance as a function of the pulse number when potentiating the device by erasing pulses with the width of 10 us.

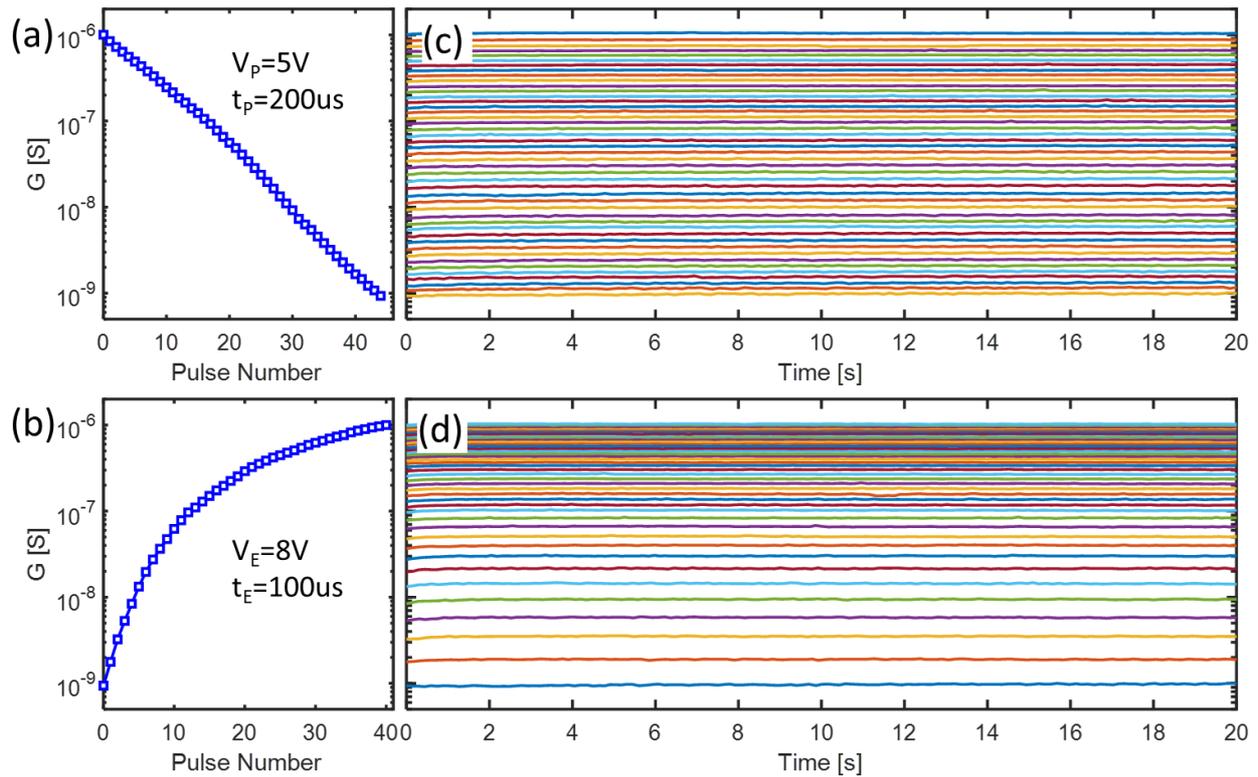

**Extended Data Fig. 3 | Depression, potentiation, and continuous readout operations of the devices.** (a) Depression (program) of the device by 200 μs pulses. (b) Potentiation (erase) of the device by 100 μs pulses. (c) The conductance of the device as a function of time after each program pulse by continuously reading the state of the device for 20 seconds. (d) The conductance of the device as a function of time after each erase pulse by continuously reading the state of the device for 20 seconds.

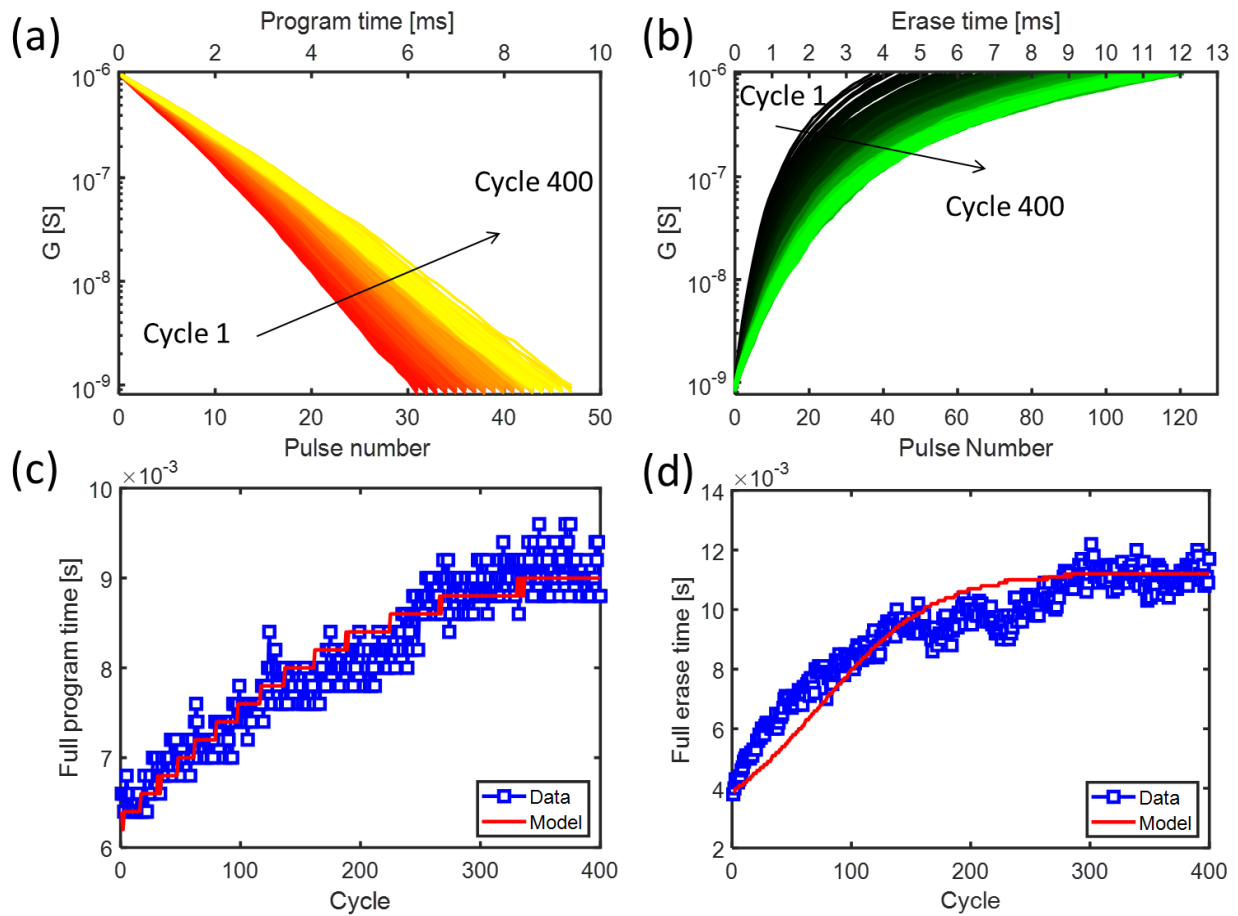

**Extended Data Fig. 4 | Cycling degradation and its modeling.** (a) Continuously depression/program of one device from the high conductance state (HCS) to the low conductance state (LCS) for 400 test cycles. (b) Continuously potentiation/erase of one device from the LCS to the HCS for 400 test cycles. (c) The program time (pulse number multiplied by the pulse width) needed for programming the device from HCS to LCS as a function of the cycling number. Both the experimental data and simulation results are presented in the figure. (d) The erase time needed for erasing the device from LCS to HCS as a function of the cycling number.

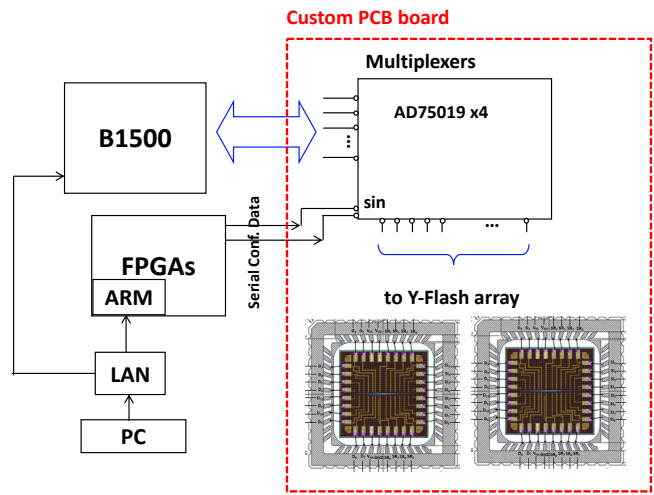

**Extended Data Fig. 5 | The schematic setup of the test system for array operations and memristive RBM demonstration.**

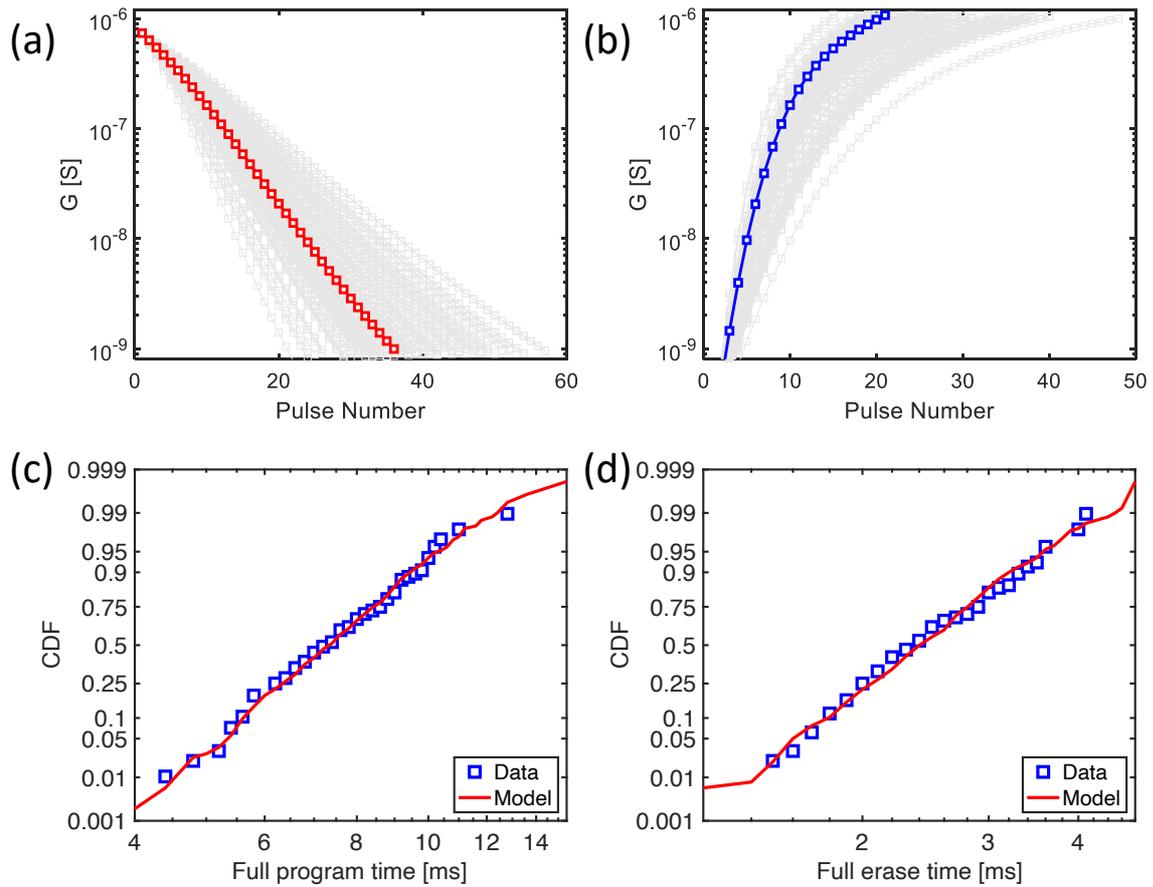

**Extended Data Fig. 6 | Modelling of device-to-device variations.** (a) Simulated device-to-device variations of the depression/program behavior (red line: a typical depression curve; gray lines: all depression curves of other devices). (b) Simulated device-to-device variations of the potentiation/erase behavior. (c) Statistical result of total program times in different devices compared with simulation results. (d) Statistical result of total erase times in different devices compared with simulation results.

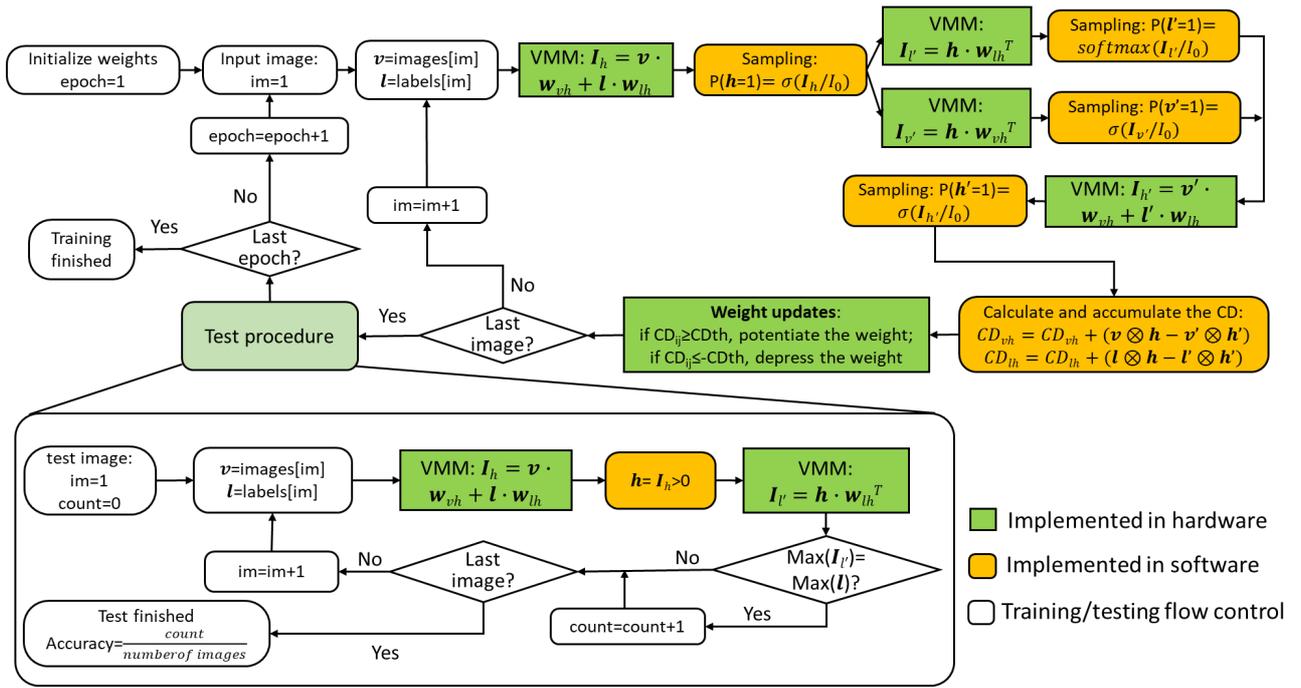

**Extended Data Fig. 7 | Flow chart of the online training of RBM including test algorithm after each training epoch.** The VMM and weight updates are performed in the memristive array using the testing system. The stochastic sampling, as well as the calculation and accumulation of the contrastive divergence (CD), are performed in software.

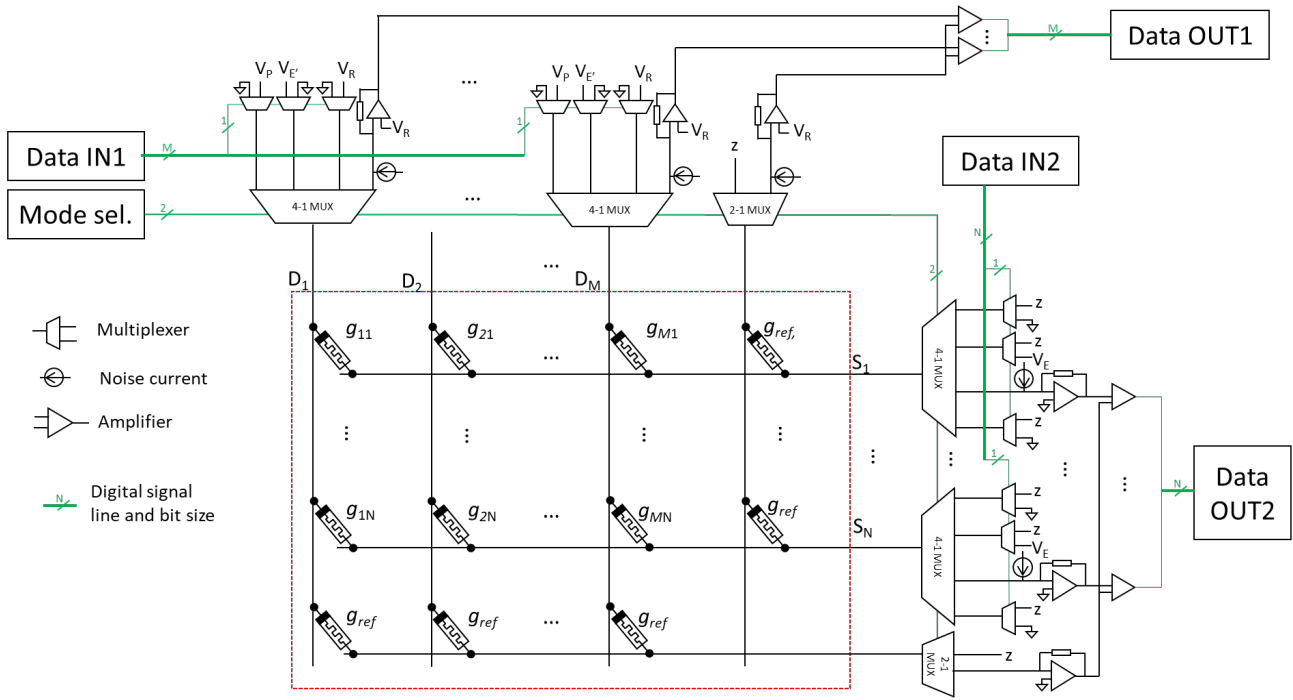

**Extended Data Fig. 8 | Full hardware design of the memristive part of an RBM layer.** The memristive part of an RBM layer conducts the forward and backward VMMs as well as the stochastic sampling in the analog domain. The peripheral circuit includes multiplexers, trans-impedance amplifiers, noise current generators, and comparators. No digital-to-analog converters (DACs) or analog-to-digital converters (ADCs) are needed in the peripheral circuit. Other parts of the memristive RBM and DBN are all digital circuits.